\documentclass{article}
\usepackage[dvips]{graphicx}
\begin{document}
\def\hf{{1 \over 2}}
\def\beq{\begin{equation}}
\def\eeq{\end{equation}}
\def\vk{\vec k}
\def\epk{\epsilon(\vec k,0,a)}
\def\epw{\epsilon_2(\vec 0,\omega,a)}
\def\ep{\epsilon}
\def\nonu{\nonumber}

%\catchline{}{}{}{}{}

\title{Metal-insulator transitions in tetrahedral semiconductors under
lattice change}

\author{Shailesh Shukla and Deepak Kumar \\
School of Physical Sciences, Jawaharlal Nehru University, \\
New Delhi-110067, India\\\\
Nitya Nath Shukla and Rajendra Prasad \\
Department of Physics, Indian Institute of Technology,\\
Kanpur-208016, India}
%\author{SHAILESH SHUKLA and DEEPAK KUMAR}

%\address{School of Physical Sciences, Jawaharlal Nehru University,
%New Delhi-110067, India}
%
%\author{NITYA NATH SHUKLA and RAJENDRA PRASAD}
%
%\address{Department of Physics, Indian Institute of Technology,
%Kanpur-208016, India}

\maketitle

\begin{abstract}

Although most insulators are expected to undergo insulator to metal transition
on lattice compression, tetrahedral semiconductors Si, GaAs and InSb can become metallic on compression
as well as by expansion. We focus on the transition by expansion which is rather
peculiar; in all cases
the direct gap at $\Gamma$ point closes on expansion and thereafter a zero-gap
state persists over a wide range of lattice
constant. The solids become metallic at an expansion of $13\%$ to $15\%$  when an
electron fermi surface around L-point and a hole fermi surface at $\Gamma
$-point develop. We provide an understanding of this behavior in terms of
arguments based on symmetry and simple tight-binding considerations.  
We also report results on the critical behavior of conductivity in the
metal phase and the static dielectric constant in the insulating phase and find
common
behaviour.  We consider the possibility of excitonic phases and distortions
which might intervene between insulating and metallic phases.

\end{abstract}

%\keywords{Metal-insulator transition, tetrahedral solids}

\newpage
\section{Introduction}
The phenomenon of metal-insulator transition in solids has been a subject of
great interest for over five decades and still continues to be a  
subject of intense research activity\cite{MI}. The phenomenon has been observed
in a large number of insulators and semiconductors as the lattice is
compressed\cite{JZ,GJA}. Thus
a basic theoretical model for studying the metal-insulator transition has been 
the study of electronic properties  of a solid whose lattice constant is
varied. 
For odd number of electrons per cell, this is a 
well-known problem first analyzed by Mott\cite{M1}. Here a metal to insulator 
transition occurs on lattice expansion due to electron-electron interactions.
For even number of electrons per cell, at a given lattice constant 
the solid can be a metal or an insulator depending on the band structure. In
this situation the change of lattice constant leads to movement of band edges,
which can lead to opening or closing of the energy gap between the valence and
conduction bands.
 Mott \cite{M2} pointed out that even for such a transition, the 
electron-hole correlations play an important role. He argued that a low density
of electrons and holes cannot exist, as in this situation the 
screening is weak and electrons and holes would pair up to form excitons.  
Over the years the conditions for formation of excitons and possibilities of 
intervening exciton-condensed phases have been studied by a number
of authors \cite{Knox}-\cite{MD}.
From these studies a complex scenario for transition from insulator to metal
phase emerges under different conditions of band closing.
 
In this work we theoretically study  metal-insulator transition in tetrahedral 
semiconductors Si, GaAs and InSb under lattice change maintaining the diamond structure. 
For these solids, the metallic phase occurs both under compression as well as expansion. 
Here we focus on the transition on expansion, where the metal phase results due to the 
closing of the hybridization gap. As is well known that at equilibrium lattice constant 
the s- and p-states mix and give rise to a hybridization gap with
four bonding s-p states below the gap and four antibonding states
above the gap, which results in an insulating state \cite{Harr}. On expansion, the s-p 
hybridization weakens and the band gap closes resulting in the atomic like ordering of bands which consists of partly filled p-like band at large atomic separations.
However, at the band picture level, a new scenario for the transition emerges. The metal 
phase does not occur 
immediately following the band closing, but requires a considerable further 
expansion. The results show that the insulating phase is separated from the
metal phase by a substantial range of lattice constant over which the system is
an insulator with zero band gap.

In this paper, we present detailed band structure calculations for the above
materials. We give simple physical arguments based on symmetry and
tight-binding considerations to understand the peculiar existence of the zero-
gap insulating phase over a rather wide range of lattice parameter. All the
calculations are done within the density functional formalism with local
density approximation\cite{DFT}. Thus the electron-hole correlation effects
mentioned above have not been incorporated in the calculation yet.

We also present  calculations of some dielectric properties. We have chosen
to examine the behavior of the imaginary part of the zero-wavevector
dielectric function, $\epsilon^{\prime\prime}(\omega)$. On the metallic side
\beq
\epsilon^{\prime\prime}(\omega) = {4 \pi \sigma^{\prime} (\omega) \over \omega}
\eeq
where $\sigma^{\prime} (\omega) $ denotes the real part of frequency dependent conductivity. 
On the insulating side, $\epsilon^{\prime\prime}(\omega)$  behaves at low frequencies
(allowing for lifetime to the electronic states) as,
\beq
\epsilon^{\prime\prime}(\omega) = A \omega
\eeq
The constant $A$ is seen to be proportional to the static dielectric
constant $\epsilon_s$ by dispersion relation. Thus on each side of the
transition, coefficients of the main $\omega$-dependent term show critical behavior.
For a continuous transition one may hope to find some universal relation
between these critical behaviors. Accordingly, on the metal side we have calculated
the dc conductivity  and find its critical behavior as the transition is approached. On
the insulating side we have calculated the band gap and the
imaginary part of the frequency-dependent dielectric function at zero
wavevector. From this we obtain the critical behaviour of the dielectric
constant.

The paper is organized as follows. In the next section, we describe the
calculational procedure for the various quantities computed here. This is
followed by a section containing results and their interpretation. Finally we  
close the paper with a discussion of the results, some speculations and concluding remarks.

\section{Calculational details}
For calculating the band structure, we use the full potential linearized
augmented plane wave(FP-LAPW) method \cite{DJS} with the WIEN CODE \cite{PB}.
The essential features of the calculational procedure are: (i) The
unit cell is divided into two parts, atomic spheres and interstitial regions.
The basis set is built by functions which are atom-like wavefunctions within
atomic spheres, and plane waves in the interstitial regions. The Kohn-Sham
orbitals of DFT are now expanded in terms
of this basis set. (ii) A modified tetrahedron method is used to do
integrations over the Brillouin zone.

These eigenenergies and eigenfunctions are used to calculate the band
parameters and response functions mentioned above. On the metallic side, we
calculate the dc conductivity, $\sigma_{dc}$, using the formula
\beq
\sigma_{dc} = {e^2 \over 3 \Gamma} \sum_{n, \vec k} v^2_{n,\vk} \delta (E_{n,\vk}
- E_F)
\eeq
where $E_{n,\vk}$ is the energy eigenvalue and $\vec v_{n,\vk} = \vec \nabla
E_{n,\vk}$, is the velocity for the eigenstate with quantum numbers ${n,\vk}$ .
$\Gamma$ denotes the transport relaxation rate, which is taken to be state-
independent here.

On the insulating side, we calculate the energy gap and the imaginary part of
the zero-wavevector dielectric constant, $\epsilon^{\prime \prime}(\omega)$ as function of 
frequency, $\omega$. For a crystalline solid, the dielectric constant is a
matrix of the form, $\epsilon_{\vec G_1 + \vec q, \vec G_2 + \vec q}(\omega)$,
where $\vec G_1$ and $\vec G_2$ are
reciprocal lattice vectors. However, here we calculate only the diagonal part
corresponding to $\vec G_1 = \vec G_2 = \vec q = \vec 0$, which amounts to
neglecting the local effects. The diagonal part is expected
to contain the more significant aspects of the critical behavior. 
For $\omega > 0$, $\epsilon^{\prime \prime}(\omega)$  is obtained from,
\beq
\epsilon^{\prime \prime}(\omega) = {4 \pi e^2 \over m^2 \omega^2} \sum_{\vk,i,f}
|<i,\vk|p_x|\vk,f>|^2 f_{i \vk}(1 - f_{f \vk})
{1 \over \pi} {\gamma \over (E_{f \vk} - E_{i \vk} - \hbar \omega)^2 +
\gamma^2}
\eeq
where $p_x$ denotes the $x$-component of momentum operator, and $f_{n,\vk}$
denotes the occupation
number of the state $n, \vk$. Note that we have allowed for a finite lifetime,
$\gamma^{-1}$, to the transition between the states \cite{FT1}, which changes the usual
$\delta$-function to the Lorentzian. Using dispersion relations, we can obtain
in principle, the real part of the dielectric function as well from this
calculation. But since our calculations are limited to a small range in
frequency, we are able to obtain information only about the static dielectric
constant.

\section{Results}
We first present the results for GaAs. For the sake of simplicity these
results do not include the effects of spin-orbit interaction(SO)
\cite{FT3,LOUIE}.
 Fig.1 shows the band structure
plots for four values of the lattice constant, $a$.
Since we are discussing
the situation at zero temperature, we take the zero
of energy to be the highest occupied level, labelled as $E_F$. Fig. 1(a) 
corresponds to equilibrium lattice constant $a = 10.68 a.u.$. 
As $a$ is increased, the direct gap begins to decrease, closing 
at $a = 10.82 a.u.$, which is
shown in panel (b). The next panel (c) shows the band structure
at $a = 11.90 a.u.$. 
One notes here that though there is considerable movement of bands,
the fermi level $E_F$
remains stuck at
$\Gamma_{15}$. This situation continues till
 $a = 12.24 a.u.$,
where the lowest conduction band touches $E_F$ at $L$ as
shown in Fig. 1(d). Thus the system persists as
 a zero-gap insulator over the lattice constant range
from $a = 10.82 a.u.$ to
 $12.24 a.u.$. Thereafter the system becomes metallic with
density of states at $E_F$
 rising rapidly. In Fig. 2, we show the plots of the density of states
(DOS) at 4 corresponding values of the lattice parameter. In Fig. 2(b)
and (c) the DOS
around $E_F$ is nonzero but too small to be seen in the figures. These plots
further confirm the zero band gap state persisting in the entire region from
$a = 10.82 a.u.$ to $12.24 a.u.$.

Fig. 3 shows band structure results for Si for
four values of lattice constant. Fig. 3(a) shows the results for the
equilibrium lattice constant $a = 10.26 a.u.$. Here the smallest band
gap is between the valence band at $\Gamma$ and the conduction band 
around X in $\Gamma$-X direction. In Fig. 3(b) at
 $a = 11.17 a.u.$, one sees that the direct gap at
$\Gamma$ becomes the smallest. With further expansion at $a = 11.39
a.u.$,
the direct gap closes as seen in Fig. 3(c). Thereafter the system
stays in the zero gap state till  $a = 11.96 a.u.$, where the conduction band at
L touches $E_F$ as seen in Fig. 3(d). The density of states for Si on lattice expansion
shows behavior similar to that of GaAs and hence is not shown here.  
The calculation confirms that the zero gap condition
occurs over the range $a = 11.39 a.u.$ to $11.96 a.u.$ in this case.

We now provide some arguments that enable us to understand this peculiar
behavior in terms of the symmetry of the
structure and the symmetry of the atomic orbitals involved in the formation of
the relevant bands. We present these arguments in
terms of tight-binding (TB) considerations \cite{Harr,CC}. 
For GaAs,
the unit cell has two atoms per cell to be denoted as $c$ and $a$, each with a
tetrahedral surrounding of the unlike atoms. For Si, the two atoms are
identical but are taken to belong to different sublattices $c$ and $a$. 
For a reasonable description of the band structure of these solids we
require a minimum of eight orbitals, one s- and three p-orbitals 
of each c and a atoms. These are labeled as: $s^c, s^a, p^c_x, p^c_y, p^c_z,
p^a_x, p^a_y$, and $p^a_z$. For details we refer to the book by Harrison\cite
{Harr}. 

 Using the TB method we can easily understand the
degeneracies of the eight bands at symmetry points like $\Gamma$, $X$, $W$
and in symmetry directions like $\Gamma - X$ etc., as due to symmetry
the tight-binding Hamiltonian matrix\cite{Harr}
is simplified and is reducible. At $\Gamma-$point, we have two s-type
levels $\Gamma_1$ and two p-type levels
$\Gamma_{15}$ each with a degeneracy of three. These eight
levels along with their degeneracies denoted by D, are given by\cite{Harr}
\begin{eqnarray}
E_s^{1,2}(\Gamma_1)
 &=& {\ep^c_s + \ep^a_s \over 2} \pm \Big [ \big ( {\ep^c_s - \ep^a_s
\over 2}
\big)^2 + 16 M_{ss}^2 \Big]^{\hf} \; D=1 \\
E_p^{1,2}(\Gamma_{15})
 &=& {\ep^c_p + \ep^a_p \over 2} \pm \Big [ \big ( {\ep^c_p - \ep^a_p
\over 2} \big )^2 + 16 M_{xx}^2 \Big ]^{\hf} \; D=3 
\end{eqnarray}
At $X$ point, we have four hybridized $sp_x$ levels and two p-type levels of
degeneracy two each corresponding to orbitals $p_y$ and $p_z$, given by
\begin{eqnarray}
E_{sp}^{1,2} &=& {\ep^c_s + \ep^a_p \over 2} \pm \Big [ \big ( {\ep^a_p -
\ep^c_s \over 2} \big )^2 + 16 M_{sp}^2 \Big ]^{\hf} \;\; D=1 \\
E_{sp}^{3,4} &=& {\ep^a_s + \ep^c_p \over 2} \pm \Big [ \big ( {\ep^c_p -
\ep^a_s \over 2} \big)^2 + 16 M_{sp}^2 \Big ]^{\hf} \;\; D=1 \\
E_p^{1,2} &=& {\ep^c_p + \ep^a_p \over 2} \pm \Big [ \big ( {\ep^c_p - \ep^a_p
\over 2} \big )^2 + 16 M_{xy}^2 \Big ]^{\hf} \;\; D=2
\end{eqnarray}
where $\ep_{s,p}^{c,a}$ denote the energies of the four atomic levels involved,
and M's denote the combinations of matrix elements between orbitals on the
neighboring atoms.
The TB parameters $\ep_{s}^c$ etc. are taken from Ref.\cite{Harr}.
Qualitatively the band structures of GaAs and Si obtained by using TB method
are similar to that shown in Figs. 1 and 3 respectively and therefore are
not shown here.

For Si, the four sp-levels become a set of two degenerate levels at X point.
Along the $\Gamma-X$ line, we have a set of four
$sp_x$ hybridized levels made up from orbitals $s^c, s^a, p^c_x$ and $p^a_x$,
which have been labelled $sp_1$, $sp_2$, $sp_3$ and $sp_4$.
The other four orbitals $p^c_y, p^c_z, p^a_y, p^a_z$ form a set of two doubly
degenerate levels labelled as $p_a$ and $p_b$. A very similar structure exists
for L point and on the
$\Gamma-L$ line. Here again the two s-orbitals hybridize with two linear
combinations of p-orbitals (the combination which points along the cube
diagonal for each a and c atoms) to give rise to a set of four sp-hybridized
levels. The remaining four p-orbitals with combinations that point in the two
perpendicular directions to the cube diagonal, then split into a set of two
doubly degenerate levels.
 Since these features of the bands are a result of basic
lattice symmetry, this degeneracy structure is seen in the full calculation 
shown in Figs. 1 and 3. Further note that the bands $sp_1$ and $sp_3$ have an
upward curvature, while the bands $sp_3$ and $sp_4$ have downward curvature
with $\bf{k}$. Similarly the bands $p_b$ and $p_a$ curve downwards and upwards
respectively.

We can now discuss the relative positioning of these eight bands as the lattice
constant is changed. It suffices to consider the ordering of levels along the
symmetry directions $\Gamma-X$ and $\Gamma-L$ and at the symmetry points
$\Gamma$, $X$ and $L$. At the equilibrium lattice constant,
two bonding sp-hybridized
bands $sp_1$ and $sp_2$ and doubly degenerate bonding p-band $p_b$ are below
$E_F$ and the remaining antibonding bands $sp_3$ and $sp_4$, and the
degenerate band $p_a$ are above $E_F$. At $\Gamma$, $sp_2$
and $sp_4$ bands become degenerate respectively to $p_b-$ and $p_a-$bands
giving rise to two three fold degenerate levels $\Gamma_{15}$. For GaAs, the
band
gap at $\Gamma$ is between triply degenerate $\Gamma_{15}$ 
level and a
$sp_3$-level $\Gamma_1$.
 For Si at equilibrium the band gap is indirect between $p_b-$
and $sp_4-$ bands, but after a stretching up to $11.17a.u.$
 the same ordering as in
GaAs happens at $\Gamma$ as seen in Fig. 3(b). For subsequent
expansion both the solids show same band movements. 

As the lattice is expanded, $\Gamma_1$ and $\Gamma_{15}$ 
levels approach each
other and at some point they touch leading to a zero gap situation at 
$\Gamma$ where one now has a four-fold degeneracy. On further expansion,
the band $sp_2$ shifts downwards, while the $sp_3$ band sticks to 
$\Gamma_{15}$ to maintain the degeneracy to three.
Since $sp_2$ band and $p_b$ bands curve downwards, while $sp_3$ curves upwards, 
four full bands stay below $E_F$ and the four full bands above. 
For some range of lattice expansion this situation persists as the $sp_3$ band
remains above $\Gamma_{15}$. Thus the system stays in the
zero-gap insulating state. During this range of lattice expansion $sp_3$ band
shows little movement around $\Gamma$ but $sp_2$ band moves down
considerably. Note the ordering of bands at $\Gamma$ has changed; both
$\Gamma_1$ levels are below $E_F$ now.
In this situation the bands have assumed
the ordering that one would expect from atomic picture, i.e. the two s-like
bands below the two p-like bands. The metallic state is reached on further
expansion when the $sp_3$ band bends downwards along $\Gamma-L$ line to cross
$E_F$. At this point an electron like fermi surface near L
and a hole like fermi surface around $\Gamma$ begin to build up
leading to a rapid increase in density of states at $E_F$.
Note that the zero-gap state results because the degeneracy at $\Gamma_{15}$
levels has to be maintained at 3 due to the symmetry.

We next give the results for InSb, whose optical gap is
rather small (0.23eV). One therefore expects  this peculiar behaviour to
occur at smaller expansions which are experimentally feasible.
Its band structure at the equilibrium lattice constant $12.24 a.u.$ is shown in Fig. 4(a). 
As in  Reference 22, we also find it to have
a vanishing gap at the $\Gamma-$point. Including spin-orbit coupling  also does
not lead to a gap \cite{Car}. 
However, the expansion leads to similar zero gap state which persists upto $a = 13.85 a.u$. 
This  is shown
in Fig. 4(b). Again the metal phase occurs when the conduction band dips to fermi
level at the L point,
where the gap in equilibrium was 1.4 eV. Curiously in Si and GaAs also, the gap at L point
has approximately the same value when the gap at $\Gamma-$point closes. In spite of the small direct 
band gap in InSb, the metallization occurs at about $13\%$ expansion and in this respect it 
is no different from Si or GaAs.
It is interesting to note that in Si the zero gap state persists for only $4\%$ expansion 
but in GaAs as well as in InSb, it lasts over 
$13\%$ expansion . This can be attributed to the nature of bonding of these solids; 
bonding in Si is purely covalent, while in GaAs and InSb it has some ionic character 
\cite {Fuj,mele} .

We have also calculated the band gaps for both GaAs and Si at lattice parameter
values close to $a_I=10.82 a.u$ and $11.39 a.u$ respectively at which their direct gaps vanish. Its plot with  $t=\frac{a_I - a}{a_I}$ 
(Fig. 5) shows that the gap variation with lattice parameter is approximately linear  in this region.

Next we present the calculation of some dielectric properties.
Figs. 6(a) and 6(b) show plots 
of $\epsilon^{\prime\prime}(\omega)$ with $\omega$ for a number
of values of $a$  close to the transition in GaAs and Si respectively. One can see that 
these plots are linear in the small frequency range($\omega\rightarrow0$) with the slopes rising as the 
transition is approached.
In Fig. 7 we present a plot of the slope $A$ of curves in Fig. 6(a), which is proportional 
to static dielectric constant $\ep_s$, with $t$.
%=\frac{(a_I - a)}{a_I}$, where $a_I$
%denotes the lattice constant where the band gap becomes zero first on expansion. 
 The static dielectric constant is seen to
exhibit power law with $t$ and diverges as $t\rightarrow0$.
\beq
\ep _{s} \propto t^{-\alpha}
\eeq
The exponent $\alpha$ is found to be 1.8$\pm 0.05$ for both Si and GaAs.

On the metal side we have calculated the static conductivity $\sigma _{dc}$ around the 
transition point and have plotted it with  $t$ in Fig.8(a) and 8(b). This data can again be fitted to a power law in $t$, where now  $t =\frac{a - a_M}{a_M}$ and $a_M$ denotes the lattice constant at which the system becomes a metal.
\beq
\sigma _{dc} \propto t^b
\eeq 
The values of the exponent are $b=2.01 \pm .09$ for GaAs and $1.80 \pm .05$ for Si. On the
basis of this calculation we cannot claim them to be equal, but the
calculations are quite limited by numerical accuracy in the regime of small
conductivities as counting of small density of states in the Brillouin zone is prone to 
errors. Interestingly these exponents are equal within numerical accuracy
to the values obtained for these exponents for CsI, which undergoes a continuous 
metal-insulator transition on lattice compression \cite{sksp}.

\section{Discussion}
It is clear from the above  analysis  that the insulator metal transition and associated  
peculiar behaviour on lattice expansion results
from the features of band structure at points $\Gamma$, X and L which
are basically governed by the symmetry of the underlying lattice and the symmetry of s 
and p orbitals.
We expect all s-p bonded tetrahedral semiconductors to show this behaviour. Further we 
emphasize
that the transition will occur irrespective of the exchange-correlation potential used, as 
it is symmetry
driven, apart from slight differences in the values of transition point lattice
parameters $a_I$ and
$a_M$ \cite{FT2,ZHL}. The zero gap state over an extended range of lattice constants is 
admittedly somewhat hypothetical, as it is based on the assumption that the diamond 
structure is maintained on expansion. With the weakening of the hybridization, it is not 
clear if the tetrahedral symmetry would be the preferred one. However, these peculiar band 
effects may underlie interesting physics arising from electron-electron
interaction and electron-lattice interaction. We discuss some of these issues below.

An important issue is the role of electron-hole interactions as mentioned in the 
introduction, which may lead to a new exciton-condensed ground states in systems
with small or zero band gaps \cite{Halp}. We first discuss the situation from the insulating
side, where Knox \cite{Knox} argued, that when the binding energy, $E_b$, of the exciton
becomes larger than the energy gap, $E_g$, there would be a spontaneous formation of
excitons. The binding energy of the exciton
is given by, $\mu / \ep_s^2$ units of Rydbergs , where $\mu$ denotes the
reduced mass of the electron hole pair in units of electron mass. When the energy gap is 
an indirect one, $\ep_s$ remains finite as the
gap closes and the $E_b$ can become larger than $E_g$ resulting in an
exciton-condensed phase which is a charge or spin density wave \cite{Cloz}-\cite{Halp}.
However, in the present case the concerned gap is a direct one, which also leads to the
divergence of $\ep_s$ due to different symmetry of the bands across the gap at the 
$\Gamma$-point. Thus $E_g$ and $E_b$ both
tend to zero as the gap closes. If one assumes $\ep_s
\propto E_g^{-\lambda}$, it is seen that the spontaneous
formation of excitons can occur only if $\lambda < 0.5$.

In the calculations described above, we have obtained the variations of both $\ep_s$ and 
$E_g$ with reduced lattice parameter, $t$.
% For both Si and GaAS, the plot of band gap with the reduced lattice 
%parameter is shown in Fig.8.This when used along with the behaviour of $\ep_s$ with $t$ 
 This yields $\lambda \approx 1.8$ for both Si and GaAs.
Thus the exciton phase does not occur in these materials according to Knox criterion.
The possibility of the existence of the exciton instability for the direct gap case has 
been examined
further by going beyond RPA \cite{Kubler}-\cite{STVR}. Inclusion of Hartree Fock corrections
\cite{Kubler} or the contribution of excitons themselves \cite{STVR} to polarisability does 
not alter the result that the exciton instability does not occur in this situation.

Though the Knox criterion for the occurrence exciton-condensed state is not satisfied here, 
we believe that the stability of the zero gap state needs
to be examined further, as due to its high polarizability it would be susceptible to other 
instabilities.
Distortions of the Jahn-Teller type that lift the degeneracy of the touching bands at 
$\Gamma-$point
are possible, but their precise nature would depend upon the electron-phonon interaction. 
The instability
towards distortion is quite evident if we examine the situation from metallic side. Here 
the electron and
hole fermi surfaces are separated by a vector $\vec Q \approx \frac{\pi}{a}(1,1,1)$. Under 
conditions of proper
nesting this state could be unstable to a distorted metal state with wavevector
$\vec Q$. This matter is
under further investigation.

We conclude by remarking that the band structure calculations on tetrahedral semiconductors 
reported here are indicative of an unusual
insulator to metal transition in which there is an intermediate zero-gap state over a 
considerable stretch
of lattice parameter. Several factors like stability of symmetry with expansion, 
electron-electron
interactions and electron phonon interactions would modify the picture given by
these
calculations. We feel that there is a good possibility of observing some of these peculiar 
features,
particularly the ones associated with the zero-gap state in InSb. Finally we also obtain 
some common dielectric behaviour in this class of solids.

It is a pleasure to thank Profs. S. C. Agarwal,  M. K. Harbola and R. C. Budhani for helpful discussions. This work was supported by the Department of Science
and Technology, New Delhi under project Nos. SP/S2/M-46/96.
and SP/S2/M-51/96.

\newpage
$\bf{FIGURE}$ $\bf{CAPTIONS}$

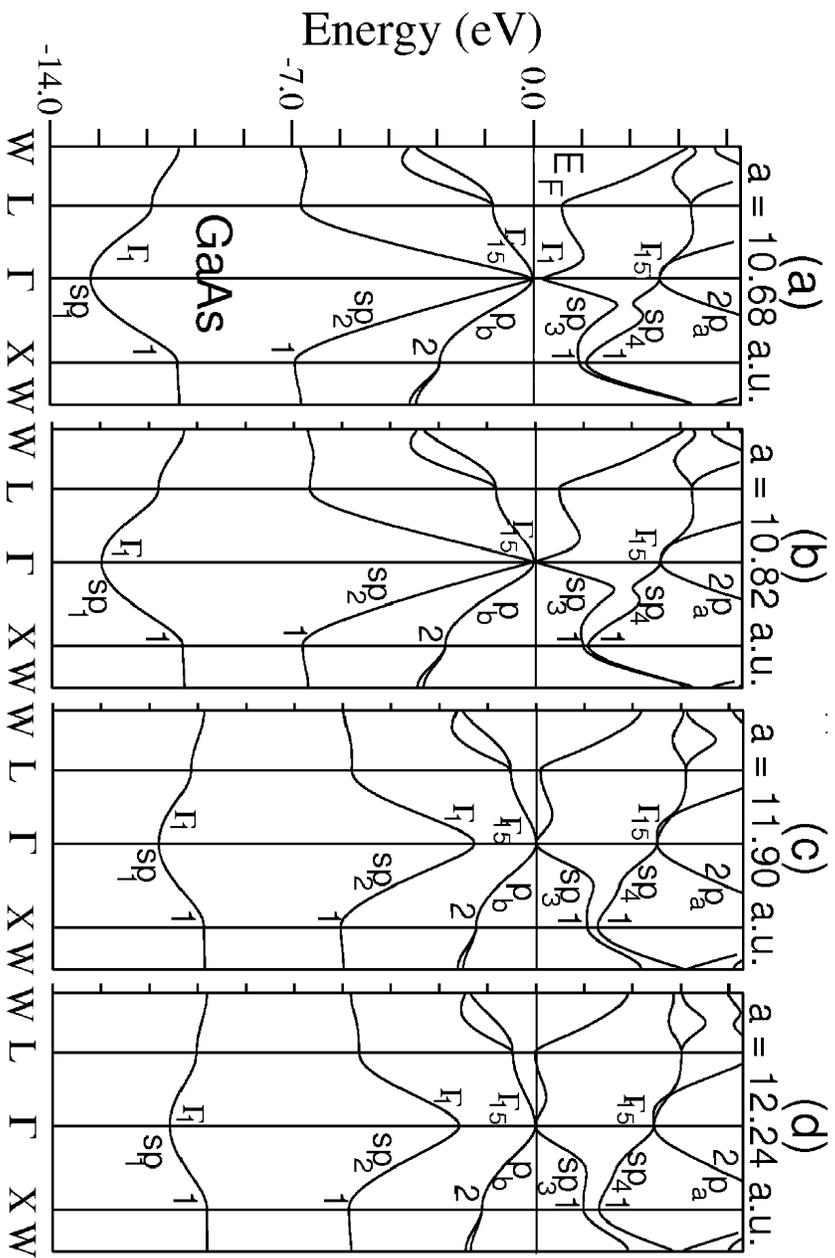
Fig. 1. Band structure of GaAs along some symmetry directions at different
lattice constants as indicated in the figure. Panel (a) corresponds to the
equilibrium lattice constant. Along $\Gamma-X$ direction, labels $sp_1$, $sp_2$,
$sp_3$ and $sp_4$
denote the sp hybridized bands and $p_a$, $p_b$, the doubly degenerate $p$ bands. Degeneracies
of the bands are denoted by numbers 1 and 2.

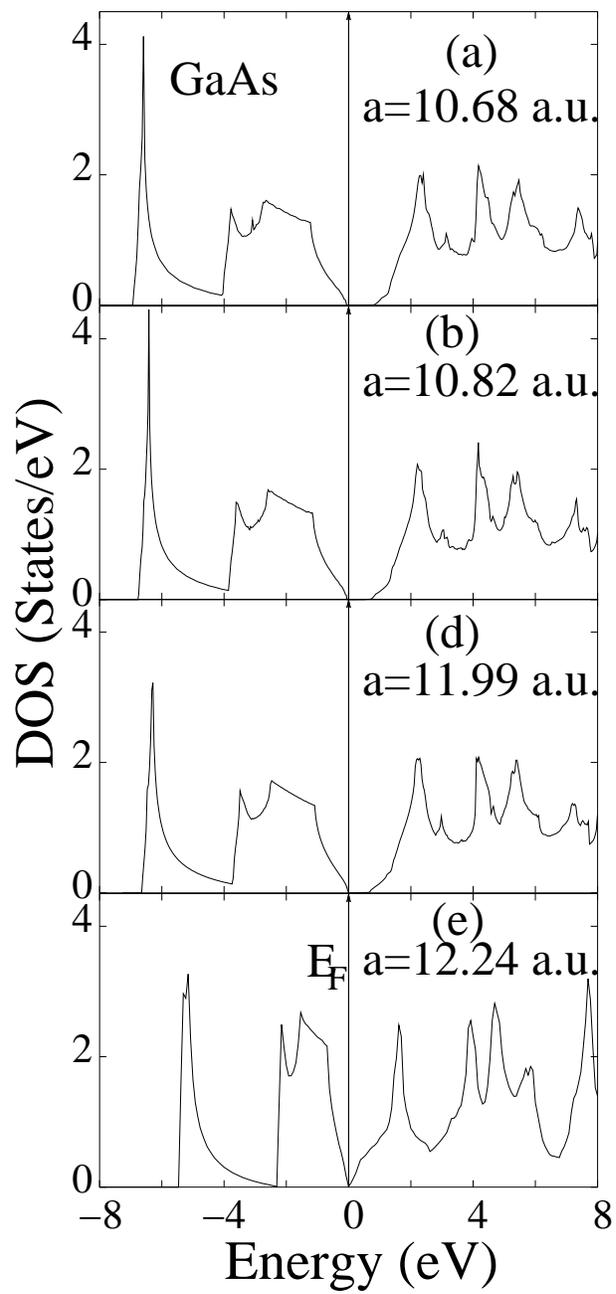
Fig. 2. Density of states for GaAs at different lattice constants as indicated
in the figure.

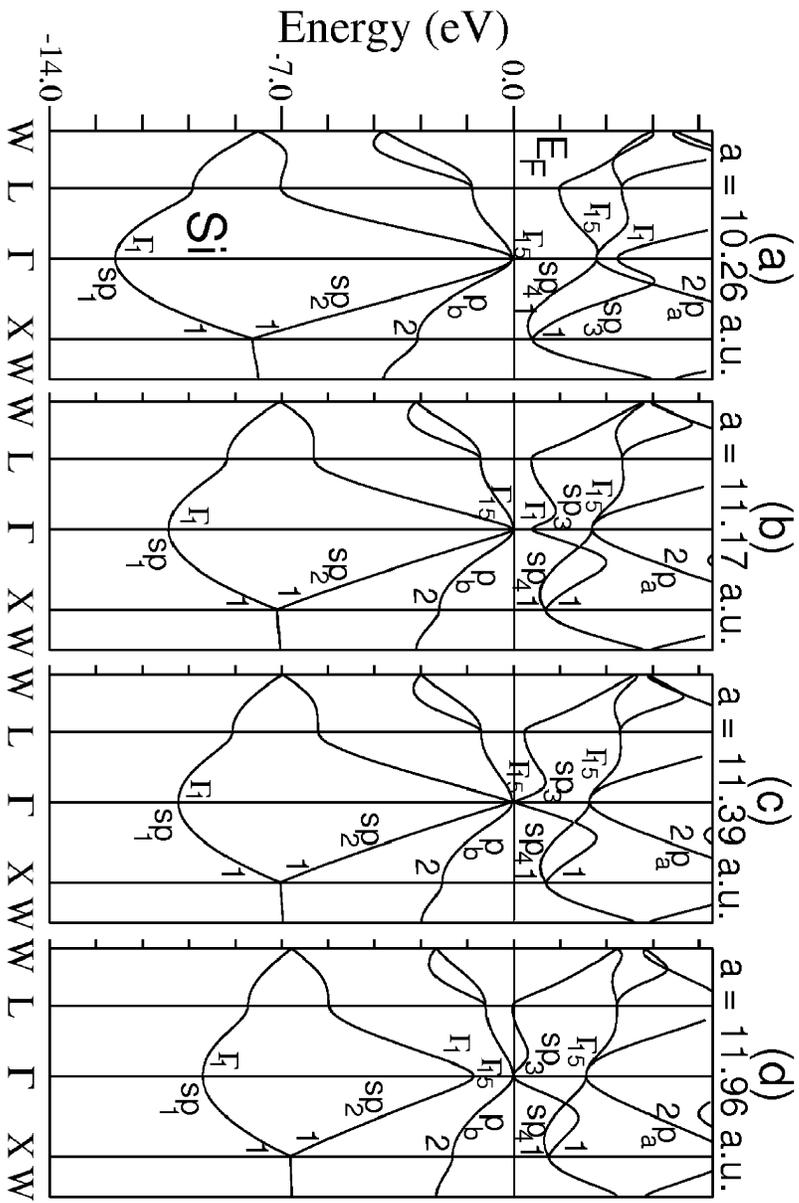
Fig. 3. Same as Fig. 1 but for Si.

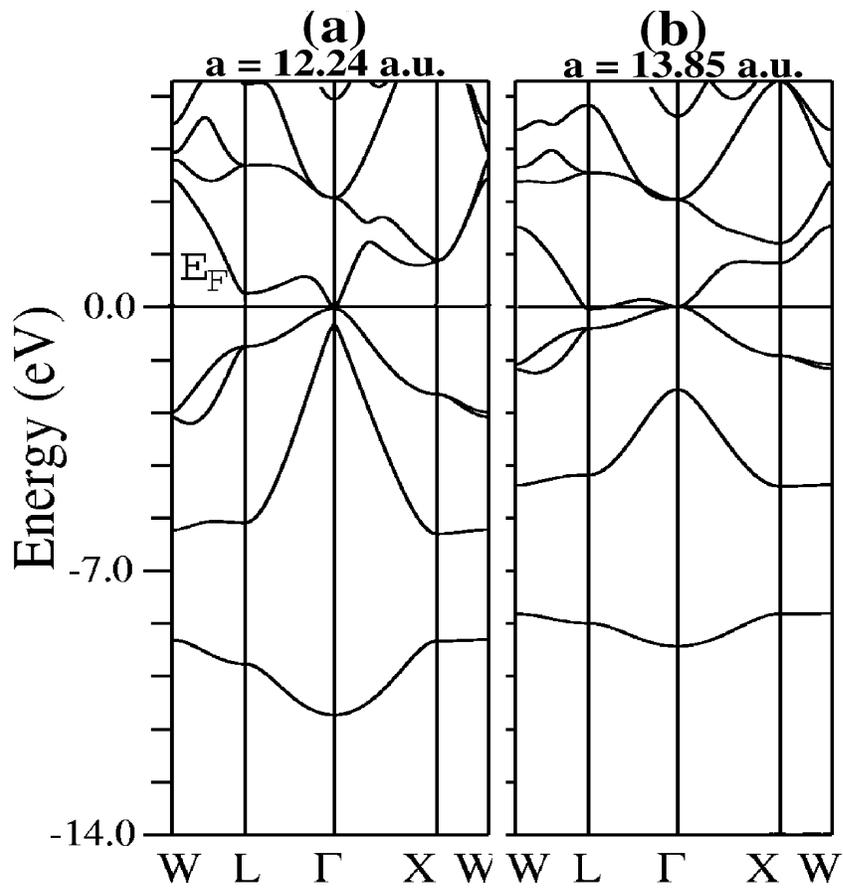
Fig. 4. Band structure of InSb at lattice constants corresponding to (a) equilibrium (b) when metallisation begins.

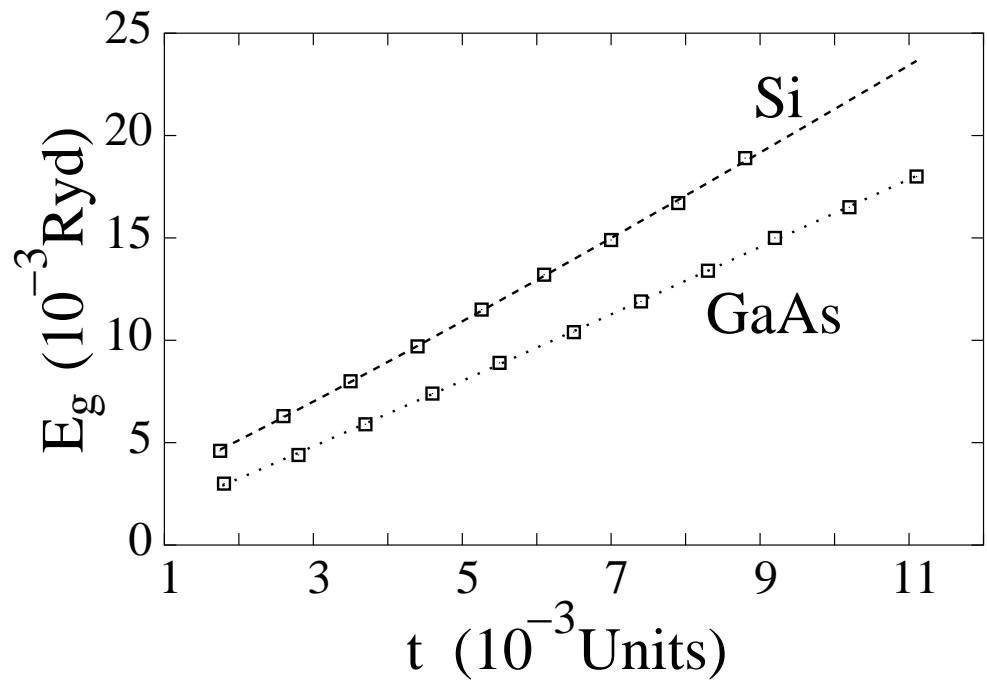
Fig. 5. Band gap variation in GaAs and Si as a function of $t=\frac{a_I-a}{a_I}$. The dashed lines show fit by $t^{1.03}$ and $t^{1.08}$ respectively.

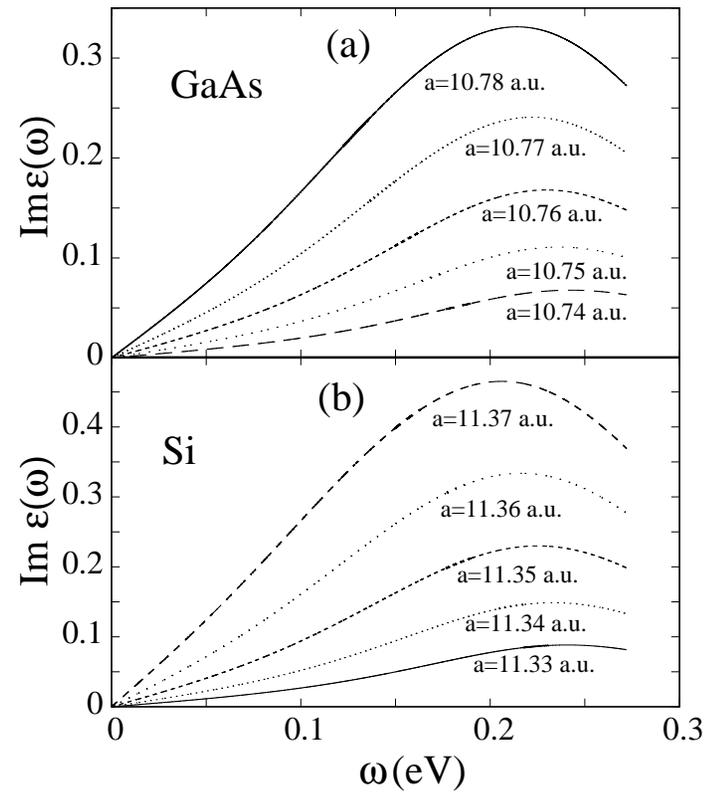
Fig. 6. (a) Imaginary part of dielectric function of GaAs versus frequency for various lattice parameters between lattice constant 10.78 a.u(top curve) and 10.74 a.u(last curve at the bottom).
 (b) Same as (a) but for Si between lattice constant 11.37 a.u(top curve) and 11.33 a.u(last curve at the bottom).

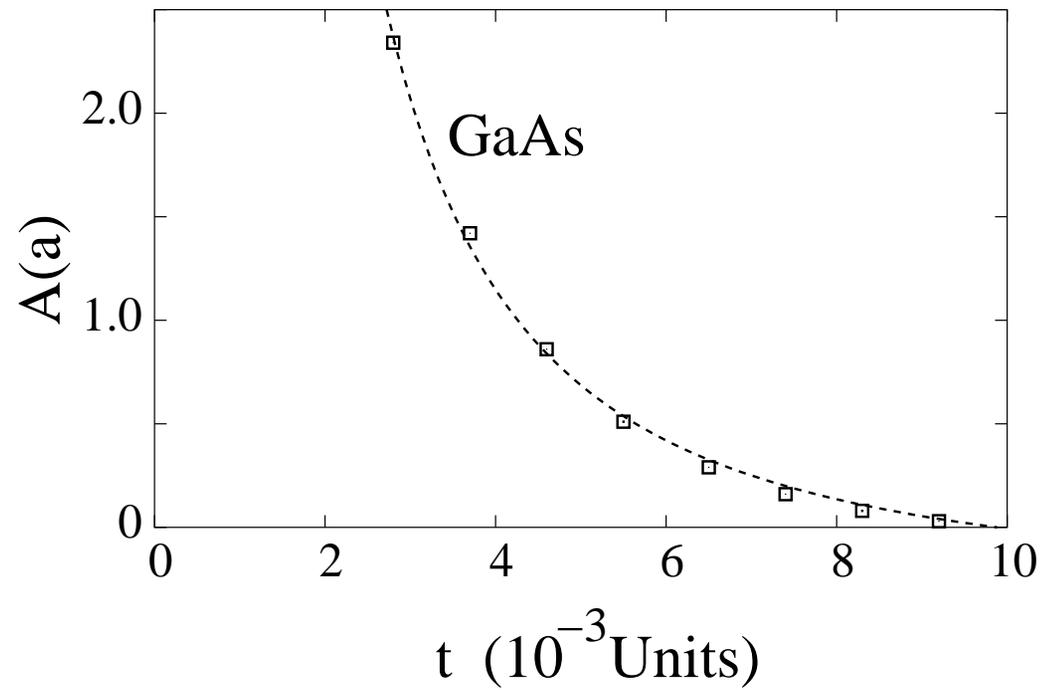
Fig. 7. Slope of the curves  shown in Fig. 6(a) in region $\omega \rightarrow 0$  as a  function of $t=\frac{a_I-a}{a_I}$. The dashed line shows power law fit by $t^{-1.79}$.

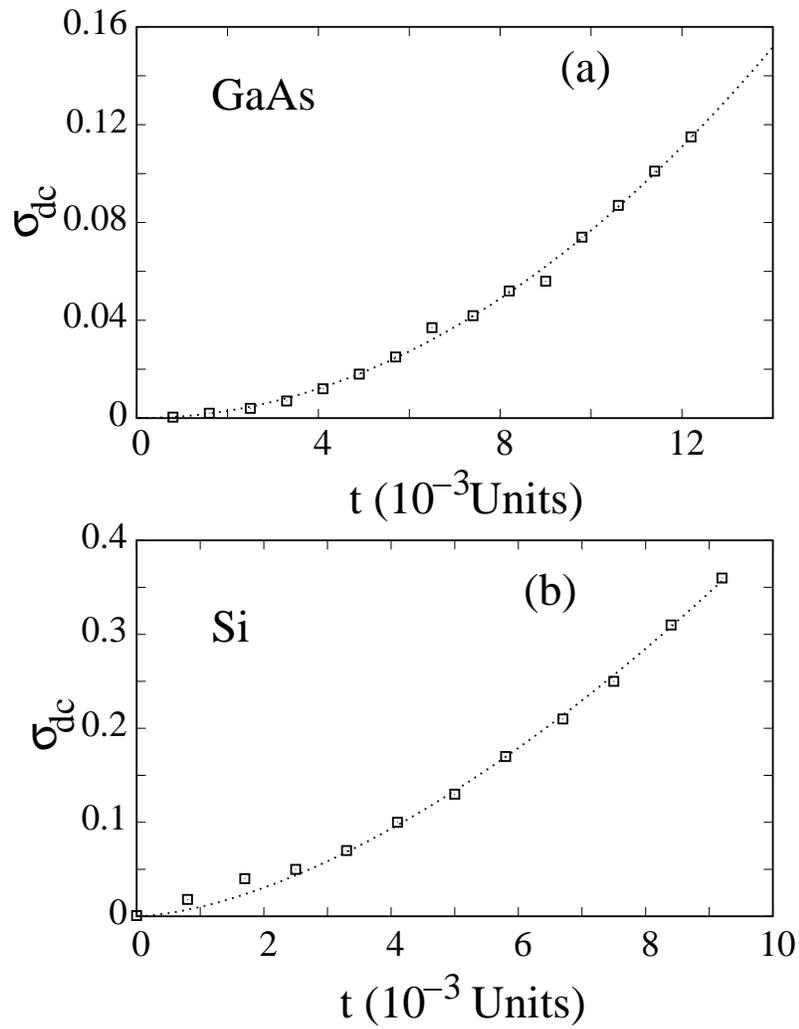
Fig. 8. (a) Plot of  d.c conductivity as a  function of $t=\frac{a-a_M}{a_M}$ for GaAs. The dashed line shows power law fit by $t^{2.01}$.
        (b) Same as (a) but for  Si. The dashed line shows power law fit by $t^{1.80}$.
\newpage
%Fig. 8. Plot of $\Gamma-$ point band gap vs reduced lattice parameter for Si
%and GaAs in expansion.
\begin{figure}[htbp]
\centering
\includegraphics[width=5.0in]{Fig1.epsi}
%\caption{Energy E of an electron in a crystal plotted as a function of the wave
%number k.}
%\label{ekfig}
\end{figure}
\begin{figure}[htbp]
\centering
\includegraphics[width=4.5in]{Fig2.epsi}
\end{figure}
\begin{figure}[htbp]
\centering
\includegraphics[width=4.8in]{Fig3.epsi}
\end{figure}
\begin{figure}[htbp]
\centering
\includegraphics[width=5.0in]{Fig4.epsi}
\end{figure}
\begin{figure}[htbp]
\centering
\includegraphics[width=4.5in]{Fig5.epsi}
\end{figure}
\begin{figure}[htbp]
\centering
\includegraphics[width=4.8in]{Fig6.epsi}
\end{figure}
\begin{figure}[htbp]
\centering
\includegraphics[width=4.5in]{Fig7.epsi}
\end{figure}
\begin{figure}[htbp]
\centering
\includegraphics[width=4.5in]{Fig8.epsi}
\end{figure}


\begin{thebibliography}{9}
\bibitem{FT1} We have broadened the energy levels by 0.1 eV.
\bibitem{FT3} The spin-orbit (SO) band structure for GaAS calculated by us at equlibrium lattice constant
agrees well with the results reported in literature \cite{LOUIE} and on expansion shows the same
trend as shown in Fig.1. (see Fig. 4) The lattice constants at which the band gap vanishes and at
which the system becomes metallic are approximately same as calculated without including SO.
\bibitem{FT2} Ref.\cite{ZHL} has shown that
 the LDA gives correct volumes at which metallization occurs
as a function of pressure.
\end{thebibliography}
\end{document}